\documentstyle[aps,12pt]{revtex}
\begin{document}
\author{Fabi\'an H. Gaioli,$^{1,2}$ Edgardo T. Garcia Alvarez,$^{1,2}$ and Diego G.
Arb\'o$^{1,3,4}$}
\title{Quantum Brownian motion. II.}
\maketitle

\begin{abstract}
{\small This paper is devoted to generalize some previous results presented
in Gaioli et al., Int. J. Theor. Phys. 36, 2167 (1997). We evaluate the
autocorrelation function of the stochastic acceleration and study the
asymptotic evolution of the mean occupation number of a harmonic oscillator
playing the role of a Brownian particle. We also analyze some deviations
from the Bose population at low temperatures and compare it with the
deviations from the exponential decay law of an unstable quantum system. }

{\small \vspace{1.0in} }

{\small $^1$Instituto de Astronom\'\i a y F\'\i sica del Espacio, C.C. 67,
Suc. 28, 1428 Buenos Aires, Argentina. }

{\small $^2$Departamento de F\'\i sica, Facultad de Ciencias Exactas y
Naturales, Universidad de Buenos Aires, 1428 Buenos Aires, Argentina. }

{\small $^3$Comisi\'on de Investigaciones Cient\'\i ficas de la Provincia de
Buenos Aires, Argentina. }

{\small $^4$Universidad Nacional de La Matanza, F. Varela 1903, 1754 San
Justo, Prov. de Buenos Aires, Argentina. }
\end{abstract}

\section{Introduction}

This work is an extension of 
some analytical results that have already been presented
in a previous paper on Brownian motion (Gaioli {\it et al}., 1997),
hereafter referred as paper I.\footnote{%
Equations numbered as (\#) in paper I will be labeled as (I.\#).} In I we
have considered a model consisting of a Brownian oscillator of frequency $%
\Omega $ linearly coupled to a bath of harmonic oscillators in the rotating
wave approximation. This model has an exact solution from which we have
studied the time evolution of the relevant physical quantities, e.g. the
mean position $\left\langle X(t)\right\rangle $ and mean population $%
\left\langle N_\Omega (t)\right\rangle $ of the Brownian oscillator. We have
shown that the equation of motion governing the evolution of $\left\langle
X\right\rangle $ is a generalized-local-in-time form of the Langevin
equation (in mean values), with time-dependent coefficients (see also Garcia
Alvarez and Gaioli, 1998). On the other hand, the equation of motion
corresponding to $\left\langle N_\Omega \right\rangle $ is a
generalized-local-in-time form of the master equation, with time-dependent
coefficients (this was anticipated in I but finally showed in Garcia Alvarez
and Gaioli, 1998).

In Section 2 we reobtain the generalized Langevin equation (I.42) but in
this case without considering it in mean values. Then a new term appears,
which plays the role of a stochastic acceleration. Evaluating the
time-dependent coefficients through a perturbative procedure up to the first
relevant order we recover the standard form of the Langevin equation, where
the coefficients are independent of time and the stochastic acceleration has
a zero-centered distribution leading to white noise in the classical limit.

In Section 3 we show that the same perturbative analysis mentioned above
performed on equation (I.33) leads to the solution of an approximated
equation for the mean population $\left\langle N_\Omega \right\rangle ,$
introduced by van Kampen in his 1992 book [equation (XVII.2.30) of van
Kampen, 1992]. Van Kampen's equation, which is the Born approximation of the
generalized master equation derived in a previous work (Garcia Alvarez and
Gaioli, 1998), makes explicit the temporal behavior of $\left\langle
N_\Omega \right\rangle :$ It decays exponentially until reaching thermal
equilibrium with the heat bath, i.e. the Bose distribution. This behavior
has been shown in the figures of paper I, obtained from the exact solution
of the model. However, a carefully evaluation of the asymptotic value of $%
\left\langle N_\Omega \right\rangle $ shows us that some deviations from the
Bose population arise. That is, at high and intermediate temperatures the
equilibrium value corresponds to the Bose distribution at the renormalized
frequency $\Omega +\delta \Omega ,$ where $\delta \Omega $ is a shift
proportional to the square of the perturbation strength, while at low
temperatures a power-law behavior is found. This deviation is related to the
long-time tail of the decay probability $P_{\Omega \Omega }(t)$ --known as
Khalfin effect (Khalfin, 1957)-- of the unstable one-particle state $\left|
\Omega \right\rangle $ [see equation (I.34)].

In Section 4 we outline our main conclusions and there are also two
appendices: Appendix A contains the details of the calculation of the
autocorrelation function of the stochastic acceleration and Appendix B
includes an explicit derivation of the exponential decay law and the Khalfin
effect.

\section{Generalized Langevin equation and the ``stochastic'' acceleration}

Let us remember that the Hamiltonian 
[equation (I.5)] of the composed system is given by 
\begin{equation}
H=\Omega \left( B^{\dagger }B+\frac 12\right) +\sum_{n=1}^N\omega _n\left(
b_n^{\dagger }b_n+\frac 12\right) +\sum_{n=1}^Ng_n\left( Bb_n^{\dagger
}+B^{\dagger }b_n\right) .  \label{ham}
\end{equation}
The position of the subsystem Brownian oscillator $X(t)=(2M\Omega )^{-\frac 1%
2}\left[ B^{\dagger }(t)+B(t)\right] $ can be rewritten, from equation
(I.26), as

\[
X(t)=a(t)X(0)+b(t)\frac{P(0)}{M\Omega }+f(t), 
\]
where 
\begin{eqnarray}
a(t) &=&\sum_{\nu =0}^N\left| \Phi _\nu \right| ^2\cos (\alpha _\nu t), 
\nonumber \\
b(t) &=&\sum_{\nu =0}^N\left| \Phi _\nu \right| ^2\sin (\alpha _\nu t), 
\nonumber \\
f(t) &=&\frac 1{\sqrt{M\Omega }}\sum_{\nu =0}^N\sum_{n=1}^N\frac{g_n}{\alpha
_\nu -\omega _n}\left| \Phi _\nu \right| ^2  \label{cel} \\
&&\times \ \left[ \sqrt{m_n\omega _n}\cos (\alpha _\nu t)x_n(0)+\frac{\sin
(\alpha _\nu t)}{\sqrt{m_n\omega _n}}p_n(0)\right] ,  \nonumber
\end{eqnarray}
and 
\[
\left| \Phi _\nu \right| ^2=\left[ 1+\sum_{n=1}^N\left( \frac{g_n}{\alpha
_\nu -\omega _n}\right) ^2\right] ^{-1}. 
\]

Since we have two constants of integration $X(0)$ and $P(0),$ and a
particular solution $f(t),$ $X(t)$ satisfies a second order differential
equation such as

\begin{equation}
\stackrel{..}{X}(t)+\Omega ^2(t)X(t)+\Gamma (t)\stackrel{.}{X}(t)=F(t),
\label{el}
\end{equation}
with the inhomogeneous term given by 
\begin{equation}
F(t)=\stackrel{..}{f}(t)+\Omega ^2(t)f(t)+\Gamma (t)\stackrel{.}{f}(t).
\label{ae}
\end{equation}
The unknown coefficients $\Omega ^2(t)$ and $\Gamma (t)$ can be easily
determined solving the linear system which results by replacing the two
independent solutions of the homogeneous equation:

\begin{eqnarray*}
\stackrel{..}{a}(t)+\Omega ^2(t)a(t)+\Gamma (t)\stackrel{.}{a}(t) &=&0, \\
&& \\
\stackrel{..}{b}(t)+\Omega ^2(t)b(t)+\Gamma (t)\stackrel{.}{b}(t) &=&0,
\end{eqnarray*}
that is

\begin{equation}
\Omega ^2(t)=\frac{\stackrel{.}{a}\text{ }\stackrel{..}{b}-\stackrel{.}{b}%
\text{ }\stackrel{..}{a}}{a\stackrel{.}{b}-b\stackrel{.}{a}},\hspace{0.5in}%
\Gamma (t)=\frac{b\stackrel{..}{a}-a\stackrel{..}{b}}{a\stackrel{.}{b}-b%
\stackrel{.}{a}}.  \label{coefic}
\end{equation}
Garcia Alvarez and Gaioli (1998) have written these coefficients in terms of
the survival amplitude (and its complex conjugate) of the state $\left|
\omega _0\right\rangle \equiv \left| \Omega \right\rangle =B^{\dagger
}\left| 0\right\rangle $: 
\[
A_{\Omega \Omega }(t)\equiv \left\langle \Omega \left| e^{-iht}\right|
\Omega \right\rangle =a(t)+ib(t),
\]
where $h=\Omega \left| \Omega \right\rangle \left\langle \Omega \right|
+\sum\limits_{n=1}^N\omega _n\left| \omega _n\right\rangle \left\langle
\omega _n\right| +\sum\limits_{n=1}^Ng_n\left( \left| \Omega \right\rangle
\left\langle \omega _n\right| +\left| \omega _n\right\rangle \left\langle
\Omega \right| \right) +C$ [equation (I.7), we are calling now the
Hamiltonian $h$ instead of $H_1,$ according to the notation of the Garcia
Alvarez and Gaioli (1998) work], and $C$ is the zero-point energy. This
survival amplitude is the cornerstone of the theory of unstable quantum
systems.

On the other hand, the mean value of the ``stochastic'' acceleration $F(t)$
vanishes since in the thermal initial distribution (I.29) [$\rho (0)=\rho
_B(0)\otimes \frac{e^{-\beta H_b}}{{\rm tr}_b\left( e^{-\beta H_b}\right) },$
where $H_b=\sum_{n=1}^N\omega _n\left( b_n^{\dagger }b_n+1/2\right) $ and tr$%
_b$ is the partial trace over the reservoir], the bath operators have
vanishing mean values,

\begin{equation}
\left\langle F(t)\right\rangle =0.  \label{fcero}
\end{equation}

Performing a perturbative expansion up to the first relevant order in the
coupling parameter we can recover the standard form of the Langevin
equation. So, in this case we have 
\begin{equation}
\stackrel{..}{X}(t)+\left( \Omega +\delta \Omega \right) ^2X(t)+\gamma 
\stackrel{.}{X}(t)=F(t),  \label{els}
\end{equation}
where $\delta \Omega $ is the shift of the frequency and $\gamma $ the
damping coefficient.

To see this we evaluate the survival amplitude up to the second order from
(Sakurai, 1995)

\begin{equation}
A_{\Omega \Omega }(t)=e^{-i\Omega t}c_{\Omega \Omega },  \label{aoo}
\end{equation}
where

\begin{equation}
c_{\Omega \Omega }=1-\sum_{k=1}^Ng_k^2\int_0^t\int_0^{t^{\prime }}e^{i\omega
_{\Omega k}t^{\prime }}e^{i\omega _{k\Omega }t^{\prime \prime }}dt^{\prime
}dt^{\prime \prime },  \label{coo}
\end{equation}
and $\omega _{\Omega k}=\Omega -\omega _k.$ Derivating equation (\ref{coo})
with respect to time, up to second order, we obtain

\[
\frac{\stackrel{.}{c}_{\Omega \Omega }}{c_{\Omega \Omega }}=1-i\sum_{k\neq
0}^Ng_k^2\left( -i\int_0^te^{i\omega _{\Omega k}\tau }d\tau \right) .
\]
Performing a long-time approximation ($t\gg \frac 1\Omega $) 
\begin{equation}
\lim_{t\rightarrow \infty }\left( -i\int_0^te^{i\alpha \tau }d\tau \right)
=\delta _{+}(\alpha )=\frac 1{\alpha +i\varepsilon }={\rm PP}\frac 1\alpha
-i\pi \delta (\alpha ),  \label{aprox}
\end{equation}
where PP denotes the principal part. Therefore the survival amplitude of the
state $\left| \Omega \right\rangle $ has a decaying exponential contribution
[cf. equation (\ref{suram4})] 
\begin{equation}
A_{\Omega \Omega }(t)=e^{-i\left( \Omega +\delta \Omega -i\frac \gamma 2%
\right) t},  \label{a}
\end{equation}
where\footnote{%
The following expressions must be understood as if a continuous limit has
been taken in such a way that summations become integrals. On the contrary,
we must replace the principal part and delta distributions by their
corresponding approximants.} 
\begin{equation}
\delta \Omega ={\rm PP}\sum\limits_{k\neq 0}^N\frac{g_k^2}{\Omega -\omega _k}%
,  \label{deltao}
\end{equation}
and 
\begin{equation}
\gamma =2\pi \sum\limits_{k\neq 0}^Ng_k^2\delta (\omega _k-\Omega ).
\label{gama}
\end{equation}

If we now use the second order value (\ref{a}) of the amplitude $A_{\Omega
\Omega }$ into equation (\ref{coefic}), a straightforward evaluation leads to

\begin{equation}
\Omega (t)=\Omega +\delta \Omega ,\hspace{0.5in}\Gamma (t)=\gamma ,
\label{cofin}
\end{equation}
so we retrieve standard expressions usually derived from the Born and
Markovian approximations. In our case the Markovian approximation is not
necessary because equation (\ref{el}) is just local in time, in contrast
with the more known integro-differential form commonly found in the
literature (Louisell, 1973; 1977; Sargent {\it et al}., 1974; Lindenberg and
West, 1984; 1990; Ford {\it et al}., 1988; Meystre and Sargent, 1991;
Cohen-Tannoudji {\it et al}., 1992; Mandel and Wolf, 1995).

The autocorrelation function of the ``stochastic'' acceleration is defined by

\[
K(t)=\frac 12\left\langle F(0)F(t)+F(t)F(0)\right\rangle . 
\]
Up to the second order, by means of a long but straightforward calculation
(see Appendix A), we obtain

\begin{equation}
K(t)=\frac 1{2M\Omega }\sum\limits_{m=1}^N\left\{ \left[ 2\left\langle
N_m(0)\right\rangle +1\right] g_m^2\left( \omega _m+\Omega \right) ^2\cos
\omega _mt\right\} .  \label{K(t)}
\end{equation}
For an initial thermal distribution for the bath oscillators, i.e. $%
2\left\langle N_m(0)\right\rangle +1=\coth \frac{\beta \omega _m}2,$ and by
taking the limit of a continuous bath (see paper I), $K(t)$ becomes

\[
K(t)=\frac 1{2M}\int_0^\infty d\omega g^2(\omega )\frac{\left( \omega
+\Omega \right) ^2}\Omega \coth \frac{\beta \omega }2\cos \omega t, 
\]
where $g^2(\omega )$ is defined as $g^2(\omega )\Delta \omega
=\sum\limits_{n\ \left| \left( \omega <\omega _n<\omega +\Delta \omega
\right) \right. \ }g_n^2$ (Ullersma, 1966; van Kampen, 1992).

Considering a coupling function $g^2\left( \omega \right) $ which is peaked%
\footnote{%
This condition is necessary in order the rotating wave approximation to be
valid (see, e.g., Gaioli, 1997).} around $\omega =\Omega $, then we can
replace $\frac 12g^2(\omega )\frac{\left( \omega +\Omega \right) ^2}\Omega $
by $\frac \omega \pi 2\pi g^2(\Omega )$ in the integrand. Taking into
account that, up to the second order [see equation (\ref{gama})], the
damping factor in the Langevin equation is $\gamma =2\pi g^2(\Omega ),$ we
finally have

\[
K(t)\approx \frac \gamma {M\pi }\int_0^\infty d\omega \hbar \omega \coth 
\frac{\beta \hbar \omega }2\cos \omega t=\frac{\gamma k_BT}M\frac d{dt}\coth
\left( \frac{\pi k_BTt}\hbar \right) . 
\]
In the classical limit ($\hbar \rightarrow 0$) it goes to the classical
stochastic, delta-correlated, autocorrelation function

\begin{equation}
{\lim_{\hbar \rightarrow 0}}K(t)=\frac{2\gamma k_BT}M\delta \left( t\right) ,
\label{ffcl}
\end{equation}
originally proposed by Langevin.

Equation (\ref{ffcl}) corresponds to instantaneous correlated fluctuations,
which leads to a Markovian process (instantaneous memory loss, since the
values of the stochastic acceleration at two different times are not
correlated). $F(t)$ is the source of noise (fluctuations), known as {\it %
white noise}. In this limit the usual Langevin equation (\ref{els}) together
with properties (\ref{fcero}) and (\ref{ffcl}) are recovered from a general
formulation. A new ingredient is that in equation (\ref{els}) the oscillator
frequency is shifted in a value $\delta \Omega $.

\section{Behavior of the mean population of the Brownian oscillator}

Van Kampen (1992) has shown that the elimination of fast microscopic
variables leads to a closed expression for the mean value of the occupation
number of the Brownian oscillator

\begin{equation}
\frac d{dt}\left\langle N_\Omega (t)\right\rangle =-\gamma \left\langle
N_\Omega (t)\right\rangle +\frac \gamma {e^{\beta \Omega }-1},  \label{vkam}
\end{equation}
where irrelevant variables get in only through the initial thermal state at
a temperature $T=1/k_B\beta .$

In what follows we use the time-dependent perturbation theory to derive
equation (\ref{vkam}) starting from equation (I.33) of I:

\begin{equation}
\left\langle N_\Omega (t)\right\rangle =P_{\Omega \Omega }(t)\left\langle
N_\Omega (0)\right\rangle +\sum_{n=1}^NP_{\Omega n}(t)\left\langle
N_n(0)\right\rangle .  \label{itt}
\end{equation}

We consider that the perturbation [interaction of equation (\ref{ham})] is
time independent and turns on at $t=0$. In this case the Dyson series for
the transition amplitude to second order can be written as (Sakurai, 1995)

\begin{equation}
A_{nm}(t)=e^{-i\omega _mt}\left(
c_{mn}^{(0)}+c_{mn}^{(1)}+c_{mn}^{(2)}+\ldots \right) ,  \label{aper}
\end{equation}
where

\begin{eqnarray}
c_{mn}^{(0)} &=&\delta _{mn},  \nonumber \\
&&  \nonumber \\
c_{mn}^{(1)} &=&-i\int_0^te^{i\omega _{mn}t^{\prime }}v_{mn}dt^{\prime },
\label{coper} \\
&&  \nonumber \\
c_{mn}^{(2)} &=&-\sum_{k=0}^N\int_0^t\int_0^{t^{\prime }}e^{i\omega
_{mk}t^{\prime }}v_{mk}e^{i\omega _{kn}t^{\prime \prime }}v_{kn}dt^{\prime
}dt^{\prime \prime },  \nonumber
\end{eqnarray}
with $v_{mn}=\left\langle \omega _m|v|\omega _n\right\rangle =g_m\delta
_{n,0}+g_n\delta _{m,0},$ except for $v_{00}=0,$ and $\omega _{mn}=\omega
_m-\omega _n.$ The evaluation of the transition probabilities to second
order gives 
\begin{equation}
P_{nm}=\delta _{nm}+\Gamma _{nm}t,  \label{beva}
\end{equation}
with

\begin{equation}
\Gamma _{nm}=2\pi v_{nm}^2\delta _t(\omega _n-\omega _m),\hspace{0.5in}{\rm %
for}\text{ }n\neq m,  \label{gamanm}
\end{equation}
and

\begin{equation}
\Gamma _{nn}=-2\pi \sum\limits_{m\neq n}v_{nm}^2\delta _t(\omega _n-\omega
_m),  \label{gaman}
\end{equation}
where $\delta _t(\alpha )\equiv \frac{\sin ^2\alpha t}{\pi \alpha ^2t}$ is a
function which approaches Dirac's delta when time goes to infinity.\footnote{%
Let $\eta $ be the width of the interaction function $g_n^2.$ Since the
function $\delta _t\left( \alpha \right) $ has a width $4\pi /t,$ in order
to behave as a delta distribution we need times such that $t\gg 4\pi /\eta .$
This is the meaning of long times.
\par
We also see that the times involved in this approximation must satisfy $t\ll
1/\gamma ,$ where $\gamma =-\Gamma _{00},$ in order equation (\ref{beva}) to
be valid. This condition and the long-time limit restrict the time-range to $%
4\pi /\eta \ll t\ll 1/\gamma .$} Introducing these results into equation (%
\ref{itt}) and taking into account that $\Gamma _{00}=\Gamma _{\Omega \Omega
}\equiv -\gamma $ and that, in this case, $P_{\Omega \Omega }\simeq 1-\gamma
t,$ we obtain

\begin{eqnarray*}
\left\langle N_\Omega (t)\right\rangle &\simeq &\left( 1-\gamma t\right)
\left\langle N_\Omega (0)\right\rangle +\gamma t\left\langle N_{\omega
_n=\Omega }(0)\right\rangle \\
&=&\left( 1-\gamma t\right) \left\langle N_\Omega (0)\right\rangle +\frac{%
\gamma t}{e^{\beta \Omega }-1},
\end{eqnarray*}
which is the second order perturbative expansion of

\begin{equation}
\left\langle N_\Omega (t)\right\rangle \simeq e^{-\gamma t}\left\langle
N_\Omega (0)\right\rangle +\left( 1-e^{-\gamma t}\right) \frac 1{e^{\beta
\Omega }-1}.  \label{vks}
\end{equation}
Equation (\ref{vks}) corresponds to the solution of equation (\ref{vkam})
[cf. equation (XVII.2.29) of van Kampen (1992)]. This is also the second
order approximation of the solution of the exact master equation of Garcia
Alvarez and Gaioli (1998).

We have to emphasize that this is a perturbative result. A full exact
calculation can be carried out in the limit of a continuous bath (as it was
taken in Section 6 of paper I). We now calculate the asymptotic long time
equilibrium value of the mean population of the Brownian particle under this
limit. We begin with equation (I.76) but setting $\omega _{\min }$ and $%
\omega _{\max }$ equal to zero and infinity respectively, so we have 
\begin{equation}
\left\langle N_\Omega (\infty )\right\rangle =\int_0^\infty d\omega \frac{%
g^2(\omega )}{\left| R_{+}^{-1}(\omega )\right| ^2}\frac 1{e^{\beta \omega
}-1}.  \label{Nom}
\end{equation}

By considering now ``no-low'' temperatures and a small coupling it is easy
to see that $g^2(\omega )\left| R_{+}^{-1}(\omega )\right| ^{-2}$ picks the
value at $\omega =\Omega +\delta \Omega $ and therefore has a $\delta
(\omega -\Omega -\delta \Omega )$ behavior. Thus equation (\ref{Nom}) leads
to the thermal equilibrium value at the shifted frequency $\Omega +\delta
\Omega ,$ i.e.

\begin{equation}
\left\langle N_\Omega (\infty )\right\rangle =\left\langle N_{\omega =\Omega
+\delta \Omega }(0)\right\rangle =\frac 1{e^{\beta (\Omega +\delta \Omega
)}-1}.  \label{boar}
\end{equation}

In the classical limit $\left\langle N_\Omega (\infty )\right\rangle \approx
\left[ \beta (\Omega +\delta \Omega )\right] ^{-1},$ which leads to the
equipartition of energy $E\approx k_BT$ (two quadratic degrees of freedom in
a one-dimensional configuration space) and the heat capacity $C_v\approx k_B$%
.

At low temperatures, deviations (an inverse power-law falloff in expectation
values) from this equilibrium distribution were already reported by some
authors (Lindenberg and West, 1984; 1990; Haake and Reibold, 1985; Joichi 
{\it et al}., 1997), since the effect of the coupling becomes
macroscopically observable. This effect is similar to the deviation from the
exponential decay law for long times, which was first described by
Khalfin (1957). In Appendix B we explicitly calculate the survival amplitude
using the resolvent method. In such a case we show that $A_{\Omega \Omega }$
reduces to [equation (\ref{suram2})]

\begin{equation}
A_{\Omega \Omega }(t)=\int_0^\infty d\omega \frac{g^2(\omega )}{\left|
R_{+}^{-1}(\omega )\right| ^2}e^{-i\omega t}.  \label{adet}
\end{equation}
Comparing equation (\ref{adet}) with the low-temperature limit of equation (%
\ref{Nom}),

\begin{equation}
\left\langle N_\Omega (\infty )\right\rangle =\int_0^\infty d\omega \frac{%
g^2(\omega )}{\left| R_{+}^{-1}(\omega )\right| ^2}e^{-\beta \omega },
\label{Nomlt}
\end{equation}
we can see that, by making the identification $\beta =it,$ both expressions
are equivalent. That is, the low-temperature regime of the mean occupation
number corresponds to the long-time behavior of the survival amplitude of
the unstable state $\left| \Omega \right\rangle .$ This is one of the deep
interrelationships between statistical properties of nonequilibrium ensembles
and the behavior of unstable quantum systems. This was possible since our
model [equation (\ref{ham})] can be decomposed by sectors of fixed number of
quanta [see equation (I.6)], which is one of the reasons for our choice of
the model.

We then use the long-time limit of the survival amplitude (Khalfin effect)
of Appendix B [equation (\ref{ekha})] in order to estimate the
low-temperature anomalous behavior of the mean population, namely 
\begin{equation}
\left\langle N_\Omega (\infty )\right\rangle \sim \left( k_BT\right) ^{n+1}.
\label{nto}
\end{equation}
The energy of the Brownian oscillator behaves like $E\sim T^{n+1}$ and so\ $%
C_v\sim T^n,$ which is an analogous result to the low-temperature behavior
of the Debye model of phonons (Huang, 1987). One of the advantages of this
kind of calculation is that equation (\ref{nto}) can be experimentally
measured and then it provides an indirect proof of the existence of the
Khalfin effect, which is very difficult to measure because of the time scale
involved. However, the other deviation from the exponential decay law, the
Zeno effect (Misra and Sudarshan, 1977), was recently measured for 
the first time
(Wilkinson {\it et al}., 1997).

We have seen that the anomalous behavior of the mean population at low
temperatures [equation (\ref{nto})] is related with the deviations of the
exponential decay law of the unstable initially prepared
state $\left| \Omega \right\rangle$ at vey long times. 
We can also see another suggesting
relation between this anomaly and a generalized statistics proposed by
Tsallis (1988) ten years ago (see also Curado and Tsallis, 1991). Let us see
the origin of this conjecture.

B\"uy\"ukkili\c c and Demirhan (1993) and B\"uy\"ukkili\c c {\it et al}.
(1995) have shown that for a set of bosons, labeled by the index $k,$ the
mean occupation number corresponding to the generalized Bose-Einstein
canonical distribution is approximately\footnote{%
The fact that this is a non-exact result was noticed by Pennini {\it et al}.
(1995).} given by

\begin{equation}
\left\langle N_k\right\rangle \simeq \frac 1{\left[ 1+(q-1)\beta \omega
_k\right] ^{1/(q-1)}-1},  \label{tsa}
\end{equation}
where $q$ is a parameter which characterizes the nonextensive nature of the
system, and thus depends on the long-range nature of interactions present in
the system. This parameter is such that for $q\rightarrow 1$ one retrieves
the standard results, i.e. the Bose-Einstein mean population. For $q\neq 1,$
in the low-temperature regime, we can neglect the terms independent of
temperature in the denominator of expression (\ref{tsa}). Then, we have

\begin{equation}
\left\langle N_k\right\rangle \sim \left( k_BT\right) ^{1/(q-1)}.
\label{buyu}
\end{equation}

Compare equation (\ref{buyu}) with equation (\ref{nto}). We see that the
equilibrium distribution reached by the Brownian oscillator resembles that
of the bosons in thermal equilibrium for a nonextensive system, according to
Tsallis' prescription. From these equations we obtain

\[
q=\frac{n+2}{n+1},
\]
a number which satisfies $1<q<2,$ since $n>0.$ Maybe the explanation of this
behavior is that the Brownian oscillator is not able to cover all accessible
quantum states (as a consequence of strong quantum correlations at low
temperatures) and then it cannot reach the most probable distribution
according to the ergodic hypothesis.

\section{Conclusions}

Along this paper the autocorrelation function of the stochastic acceleration
and the asymptotic mean population of the Brownian oscillator were
analytically evaluated from a deterministic quantum dynamics. As regards the
Langevin equation we have provided the stochastic term which was skipped in
paper I. For the mean occupation number we have found that it reaches
thermal equilibrium at the bath temperature, corresponding to the Bose
population. At low temperatures a deviation from this population was found,
which has a common origin with the deviations from the exponential decay
law. However, the Khalfin effect is very difficult to measure since usual
observation times of unstable quantum systems are much shorter than the time
the decay law is no longer exponential.

\section*{Acknowledgments}

F.H.G. is grateful to OLAM Foundation, Mme. Smet, and the Foyer d'Humanisme
for their warm hospitality in Peyresq.

\appendix 

\section{Autocorrelation function of the stochastic acceleration}

According to equation (\ref{ae}) $K(t)$ is given by

\begin{eqnarray}
K(t) &=&\frac 12\left[ \left\langle \stackrel{..}{f}(0)\stackrel{..}{f}(t)+%
\stackrel{..}{f}(0)\Omega ^2(t)f(t)+\stackrel{..}{f}(0)\Gamma (t)\stackrel{.%
}{f}(t)\right. \right.  \label{kdt} \\
&&\ +\Omega ^2(0)f(0)\stackrel{..}{f}(t)+\Omega ^2(0)f(0)\Omega
^2(t)f(t)+\Omega ^2(0)f(0)\Gamma (t)\stackrel{.}{f}(t)  \nonumber \\
&&\ \left. +\Gamma (0)\stackrel{.}{f}(0)\stackrel{..}{f}(t)+\Gamma (0)%
\stackrel{.}{f}(0)\Omega ^2(t)f(t)+\Gamma (0)\stackrel{.}{f}(0)\Gamma (t)%
\stackrel{.}{f}(t)\right\rangle  \nonumber \\
&&\ \left. +\left\langle 0\leftrightarrow t\right\rangle \right] ,  \nonumber
\end{eqnarray}
where $\left\langle 0\leftrightarrow t\right\rangle $ stands for
interchanging $t=0$ with $t.$ $f(t)$ can be written in terms of the $%
A_{\Omega m}$'s as

\begin{equation}
f(t)=\frac 1{\sqrt{2M\Omega }}\sum_{m=1}^N\left[ A_{\Omega m}(t)b_m^{\dagger
}(0)+h.c.\right] .  \label{f}
\end{equation}

Considering second order contributions only and taking into account that $%
\delta \Omega $ and $\Gamma $ are time independent up to this order, we can
rewrite equation (\ref{kdt}) as

\begin{eqnarray*}
K(t) &=&\frac 12\left[ \left\langle \stackrel{..}{f}(0)\stackrel{..}{f}%
(t)+\Omega ^2\stackrel{..}{f}(0)f(t)+\Omega ^2f(0)\stackrel{..}{f}(t)+\Omega
^4f(0)f(t)\right\rangle \right. \\
&&\ \ \left. +\left\langle 0\leftrightarrow t\right\rangle \right] ,
\end{eqnarray*}
since $f$'s are linear in $A_{\Omega m}$'s [see equation (\ref{f})]. We need
to expand the amplitudes up to the first order. From equations (\ref{aper})
and (\ref{coper}) we have

\[
A_{\Omega m}(t)=-ie^{-i\omega _mt}v_{m\Omega }\int_0^te^{i\left( \omega
_m-\Omega \right) t^{\prime }}dt^{\prime }. 
\]
Solving the integral it is

\[
A_{\Omega m}(t)=\frac{v_{m\Omega }}{\omega _m-\Omega }\left( e^{-i\omega
_mt}-e^{-i\Omega t}\right) . 
\]
The second derivative of $A_{\Omega m}(t),$ appearing in $\stackrel{..}{f},$
is given by

\[
\stackrel{..}{A}_{\Omega m}(t)=\frac{v_{m\Omega }}{\omega _m-\Omega }\left(
-\omega _m^2e^{-i\omega _mt}+\Omega ^2e^{-i\Omega t}\right) . 
\]
Taking into account that $A_{\Omega m}(0)=0,$ from which $f(0)=0,$ then we
must only calculate

\[
K(t)=\frac 12\left[ \left\langle \stackrel{..}{f}(0)\stackrel{..}{f}%
(t)+\Omega ^2\stackrel{..}{f}(0)f(t)\right\rangle +\left\langle
0\leftrightarrow t\right\rangle \right] . 
\]
Let us see each term step by step. Let $K_1$ and $K_2$ be defined by

\[
K_1(t)=\frac 12\left\langle \stackrel{..}{f}(0)\stackrel{..}{f}(t)+\stackrel{%
..}{f}(t)\stackrel{..}{f}(0)\right\rangle 
\]
and

\[
K_2(t)=\frac{\Omega ^2}2\left\langle \stackrel{..}{f}(0)f(t)+f(t)\stackrel{..%
}{f}(0)\right\rangle . 
\]
Therefore

\begin{eqnarray*}
K_1(t) &=&\frac 12\frac 1{2M\Omega }\sum\limits_{m,m^{\prime }=1}^N\left[ 
\stackrel{..}{A}_{\Omega m}(0)\stackrel{..}{A}_{\Omega m^{\prime
}}^{*}(t)\left\langle b_m^{\dagger }b_{m^{\prime }}\right\rangle (0)+%
\stackrel{..}{A}_{\Omega m}^{*}(0)\stackrel{..}{A}_{\Omega m^{\prime
}}(t)\left\langle b_mb_{m^{\prime }}^{\dagger }\right\rangle (0)\right. \\
&&\ \left. +\stackrel{..}{A}_{\Omega m}(t)\stackrel{..}{A}_{\Omega m^{\prime
}}^{*}(0)\left\langle b_m^{\dagger }b_{m^{\prime }}\right\rangle (0)+%
\stackrel{..}{A}_{\Omega m}^{*}(t)\stackrel{..}{A}_{\Omega m^{\prime
}}(0)\left\langle b_mb_{m^{\prime }}^{\dagger }\right\rangle (0)\right] ,
\end{eqnarray*}
which, from equation (I.30) and the commutation relations (I.4), is reduced
to

\begin{eqnarray*}
K_1(t) &=&\frac 1{2M\Omega }\sum\limits_{m=1}^N\left\{ {\rm Re}\left[ 
\stackrel{..}{A}_{\Omega m}(0)\stackrel{..}{A}_{\Omega m}^{*}(t)\right]
\left[ 2\left\langle N_m\right\rangle (0)+1\right] \right\} \\
&=&\frac 1{2M\Omega }\sum\limits_{m=1}^N\left\{ \left[ 2\left\langle
N_m\right\rangle (0)+1\right] \frac{\left| v_{m\Omega }\right| ^2}{\omega
_m-\Omega }\left( \omega _m+\Omega \right) \left( \omega _m^2\cos \omega
_mt-\Omega ^2\cos \Omega t\right) \right\} .
\end{eqnarray*}
An analogous calculation for $K_2(t)$ leads to

\begin{eqnarray*}
K_2(t) &=&\frac \Omega {2M}\sum\limits_{m=1}^N\left\{ {\rm Re}\left[ 
\stackrel{..}{A}_{\Omega m}(0)A_{\Omega m}^{*}(t)\right] \left[
2\left\langle N_m\right\rangle (0)+1\right] \right\} \\
\ &=&\frac 1{2M\Omega }\sum\limits_{m=1}^N\left\{ \left[ 2\left\langle
N_m\right\rangle (0)+1\right] \frac{\left| v_{m\Omega }\right| ^2}{\omega
_m-\Omega }\left( \omega _m+\Omega \right) \left( \Omega ^2\cos \Omega
t-\Omega ^2\cos \omega _mt\right) \right\} .
\end{eqnarray*}
Joining $K_1$ and $K_2$ we finally obtain equation (\ref{K(t)})

\[
K(t)=\frac 1{2M\Omega }\sum\limits_{m=1}^N\left\{ \left[ 2\left\langle
N_m\right\rangle (0)+1\right] \left| v_{m\Omega }\right| ^2\left( \omega
_m+\Omega \right) ^2\cos \omega _mt\right\} . 
\]

\section{Evaluation of the survival and transition amplitudes}

We analyze the analytical structure of $A_{\Omega \Omega }(t)$ in order to
show how an exponential contribution arises for a significant range of time
and how deviations from this behavior appear. Using the well-known identity
between distributions 
\begin{equation}
\frac 1{x\pm i\epsilon }={\rm PV}\frac 1x\mp i\pi \delta (x),  \label{ppipi}
\end{equation}
we can obtain, for $x=\omega -h,$ the following integral representation of
the evolution operator

\begin{equation}
e^{-iht}=\frac 1{2\pi i}\int_{-\infty }^\infty d\omega e^{-i\omega t}\left[
G_{-}(\omega )-G_{+}(\omega )\right] ,  \label{intr}
\end{equation}
where $G_{\pm }(\alpha )=\frac 1{\alpha \pm i\epsilon -h}$ is the retarded ($%
+$) $\left[ \text{advanced }(-)\right] $ Green function of the time
independent Schr\"odinger equation (the resolvent), corresponding to the
total Hamiltonian. From (\ref{intr}) we can evaluate the survival and
transition amplitudes. Such calculation involves the knowledge of the
partial resolvents (Schwinger, 1961; Messiah, 1962), departing from 
\begin{equation}
(\alpha \pm i\epsilon -h)G_{\pm }(\alpha )=I.  \label{me}
\end{equation}
By taking the matrix elements in equation (\ref{me}) we have

\begin{equation}
(\alpha \pm i\epsilon -\Omega )\left\langle \Omega \right| G_{\pm }(\alpha
)\left| \Omega \right\rangle +\int_0^\infty d\omega g(\omega )\left\langle
\omega \right| G_{\pm }(\alpha )\left| \Omega \right\rangle =1,  \label{pr1}
\end{equation}

\begin{equation}
(\alpha \pm i\epsilon -\omega )\left\langle \omega \right| G_{\pm }(\alpha
)\left| \Omega \right\rangle =g(\omega )\left\langle \Omega \right| G_{\pm
}(\alpha )\left| \Omega \right\rangle =0.  \label{pr2}
\end{equation}
So, from (\ref{pr1}) and (\ref{pr2}), the desired matrix elements of the
resolvent are

\[
\left\langle \Omega \right| G_{\pm }(\alpha )\left| \Omega \right\rangle =%
\frac 1{\alpha \pm i\epsilon -\Omega -\int_0^\infty d\omega \frac{g^2(\omega
)}{\alpha \pm i\epsilon -\omega }}=R_{\pm }(\alpha ) 
\]
and

\[
\left\langle \omega \right| G_{\pm }(\alpha )\left| \Omega \right\rangle =%
\frac{g(\omega )}{\alpha \pm i\epsilon -\omega }\left\langle \Omega \right|
G_{\pm }(\alpha )\left| \Omega \right\rangle . 
\]
Now returning to (\ref{intr}) we can obtain the survival and transition
amplitudes as

\begin{equation}
A_{\Omega \Omega }(t)=\left\langle \Omega \right| e^{-iht}\left| \Omega
\right\rangle =\frac 1{2\pi i}\int_{-\infty }^\infty d\omega ^{\prime
}e^{-i\omega ^{\prime }t}\left[ R_{-}(\omega ^{\prime })-R_{+}(\omega
^{\prime })\right] ,  \label{suramp1}
\end{equation}

\begin{equation}
A_{\Omega \omega }(t)=\left\langle \omega \right| e^{-iht}\left| \Omega
\right\rangle =\frac 1{2\pi i}\int_{-\infty }^\infty d\omega ^{\prime
}e^{-i\omega ^{\prime }t}\left[ \frac{g(\omega ^{\prime })R_{-}(\omega
^{\prime })}{\omega -i\epsilon -\omega ^{\prime }}-\frac{g(\omega ^{\prime
})R_{+}(\omega ^{\prime })}{\omega +i\epsilon -\omega ^{\prime }}\right] ,
\label{nsc}
\end{equation}
i.e. taking the Fourier transform of the difference between the advanced and
retarded reduced resolvents (note that we have not introduced any initial
condition for the amplitudes). It can be proved that the reduced resolvents
have the general expression 
\[
R_{\pm }(\omega )=\frac 1{\omega \pm i\epsilon -\Omega -\Sigma _{\pm
}(\omega )},
\]
where $\Sigma (\omega )$ is the level-shift operator in the subspace
generated by $\left| \Omega \right\rangle .$ The reduced resolvent $%
R_{+}(\omega )$ [$R_{-}(\omega )]$ is the analogous of the exact Feynman
(Dyson) electron propagator with $\Sigma _{\pm }(\omega )$ playing the role
of the Dyson (1949) mass operator. In our case $\Sigma _{\pm }(\alpha
)=\int_0^\infty d\omega \frac{g^2(\omega )}{\alpha \pm i\epsilon -\omega }$.
Using (\ref{ppipi}) we can rewrite it as $\Sigma _{\pm }(\alpha )=\Delta
(\alpha )\mp i\frac{\gamma (\alpha )}2,$ with

\begin{eqnarray*}
\Delta (\alpha ) &=&{\rm PV}\int_0^\infty d\omega \frac{g^2(\omega )}{\alpha
-\omega }, \\
&& \\
\gamma (\alpha ) &=&2\pi g^2(\alpha ),
\end{eqnarray*}
which are nothing else than equations (\ref{deltao}) and (\ref{gama}) in the
case of a continuous bath. Taking into account that $g(\omega )=0$ for $%
\omega <0,$ equation (\ref{suramp1}) can be rewritten as

\begin{equation}
A_{\Omega \Omega }(t)=\int_0^\infty d\omega \frac{g^2(\omega )}{\left|
R_{+}^{-1}(\omega )\right| ^2}e^{-i\omega t}.  \label{suram2}
\end{equation}
In the theory of unstable states (Messiah, 1962; Goldberger and Watson,
1964; Cohen-Tannoudji {\it et al.}, 1992) it is common to find $A_{\Omega
\Omega }$ written as 
\begin{equation}
A_{\Omega \Omega }(t)=\frac 1{2\pi }\int\limits_{-\infty }^\infty d\omega 
\frac{\gamma (\omega )}{\left[ \omega -\Omega -\Delta (\omega )\right] ^2+%
\frac 14\gamma ^2(\omega )}e^{-i\omega t},  \label{suram3}
\end{equation}
which allows one to easily study the different decay regimes. If $\gamma
(\omega )$ is small, the term $\left[ \omega -\Omega -\Delta (\omega
)\right] ^2$ is large compared with $\gamma ^2(\omega )$ except when $\omega
\simeq \Omega +\Delta (\Omega ).$ Thus we replace $\gamma (\omega )$ by $%
\gamma \left( \Omega \right) \equiv \gamma $ and $\Delta (\omega )$ by $%
\Delta (\Omega )\equiv \delta \Omega .$ The Lorentzian function resulting
from this replacement is known as Breit-Wigner (1936) distribution. In this
case equation (\ref{suram3}) has an analytical result 
\begin{equation}
A_{\Omega \Omega }(t)=e^{-i(\Omega +\delta \Omega )t}e^{-\frac \gamma 2t},
\label{suram4}
\end{equation}
which, as it is expected, retrieves the well-known exponential decay law, as
it was originally derived by Weisskopf and Wigner (1930). Some deviations
from this exponential decay arise as we inspect equation (\ref{suram3}) more
carefully. If we retain $\gamma (\omega )$ in the numerator of (\ref{suram3}%
), we can rewrite this equation as 
\begin{equation}
A_{\Omega \Omega }(t)=\frac 1{2\pi }\int\limits_{-\infty }^\infty d\omega 
\frac \gamma {\left[ \omega -\Omega -\delta \Omega \right] ^2+\frac 14\gamma
^2}\frac{\gamma (\omega )}\gamma e^{-i\omega t}.  \label{rhs}
\end{equation}
The r.h.s. of equation (\ref{rhs}) is the convolution product of (\ref
{suram4}) and the Fourier transform of $\gamma ^{-1}\gamma (\omega ).$ Since 
$\gamma (\omega )$ has a finite width, its Fourier transform $\gamma (t)$
has also a finite width. We also have that $\gamma (\omega )$ is null for $%
\omega \leq 0$ and is not infinitely differentiable at $\omega =0$. Then, if
we suppose that $\gamma (\omega )$ goes as $\omega ^n$ for small $\omega ,$
then $\gamma (t)$ behaves like $t^{-(n+1)}$ for very long times. This is
known as Khalfin (1957) effect, namely

\begin{equation}
A_{\Omega \Omega }(t)\sim \frac 1{t^{n+1}}.  \label{ekha}
\end{equation}

\medskip

\noindent 
{\large {\bf REFERENCES}}

{\small \noindent
Breit, G., and Wigner, E.P. (1936). {\it Physical Review,} {\bf 49}, 519. }

{\small \noindent
B\"uy\"ukkili\c c, F., and Demirhan, D. (1993). {\it Physics Letters A}, 
{\bf 181}, 24. }

{\small \noindent
B\"uy\"ukkili\c c, F., Demirhan, D., and G\"ule\c c, A. (1995). {\it Physics
Letters A}, {\bf 197}, 209. }

{\small \noindent
Cohen-Tannoudji, C., Dupont-Roc, J. and Grynberg, G. (1992),{\it \
Atom-Photon Interactions (Basic Processes and Applications)}, Wiley, New
York. }

{\small \noindent
Curado, E.M.F., and Tsallis, C. (1991). {\it Journal of Physics A}, {\bf 24}%
, L69; {\it corrigenda}: {\bf 24}, 3187 (1991); {\bf 25}, 1019 (1992). }

{\small \noindent
Dyson, F.J. (1949). {\it Physical Review}, {\bf 75}, 486; 1736. }

{\small \noindent
Ford, G.W., Lewis, J.T., and O'Connell, R.F. (1988). {\it Journal of
Statistical Physics}, {\bf 53}, 439; (1988). {\it Physical Review A,} {\bf 37%
}, 4419. }

{\small \noindent
Gaioli, F.H. (1997). {\sl Dissipation in quantum Brownian motion}, {\it %
Ph.D. Thesis}, University of Buenos Aires. }

{\small \noindent
Gaioli, F.H., Garcia Alvarez, E.T., and Guevara, J. (1997). {\it %
International Journal of Theoretical Physics}, {\bf 36}, 2167. }

{\small \noindent
Garcia Alvarez, E.T., and Gaioli, F.H. (1998). {\sl Exact derivation of the
Langevin and master equations for harmonic quantum Brownian motion}, {\it %
Physica A} (in press). }

{\small \noindent
Goldberger, M.L., and Watson, K.M. (1964). {\it Collision Theory, }Wiley,
New York. }

{\small \noindent
Haake, F., and Reibold, R. (1985). {\it Physical Review A,} {\bf 32}, 2462. }

{\small \noindent
Huang, K. (1963). {\it Statistical Mechanics, }Wiley, New York. }

{\small \noindent
Joichi, I., Matsumoto, S., and Yoshimura, M. (1997). {\it Progress of
Theoretical Physics}, {\bf 98}, 9. }

{\small \noindent
Khalfin, L. (1957). {\it Zhurnal Eksperimental'noi i Teoreticheskoi Fiziki}, 
{\bf 33}, 1371 [{\it Soviet Physics, JEPT}, {\bf 6}, 1053 (1958)]. }

{\small \noindent
Lindenberg, K., and West, B.J. (1984). {\it Physical Review A}, {\bf 30},
568; (1990). {\it The Nonequilibrium Statistical Mechanics of Open and
Closed Systems}, VCH Publishers, New York. }

{\small \noindent
Louisell, W.H. (1973). {\it Quantum Statistical Properties of Radiation},
Wiley, New York; (1977). {\it Radiation and Noise in Quantum Electronics},
Krieger, New York. }

{\small \noindent
Mandel, L., and Wolf, E. (1995). {\it Optical Coherence and Quantum Optics,}
Springer, Berlin. }

{\small \noindent
Messiah, A. (1964). {\it M\'ecanique Quantique, }Vol. 2,{\it \ }Dunod,
Paris, Chap. XXI-13. }

{\small \noindent
Meystre, P., and Sargent, M. (1991). {\it Elements of Quantum Optics},
Springer, Berlin. }

{\small \noindent
Misra, B., and Sudarshan, E.C.G. (1977). {\it Journal of Mathematical
Physics,} {\bf 18}, 756. }

{\small \noindent
Pennini, F., Plastino, A., and Plastino, A.R. (1995). {\it Physics Letters A}%
, {\bf 208}, 309. }

{\small \noindent
Sakurai, J.J. (1995). {\it Modern Quantum Mechanics}, Addison-Wesley, New
York. }

{\small \noindent
Sargent, M., Scully, M.O., and Lamb, W.E. (1974). {\it Laser Physics},
Addison-Wesley, New York. }

{\small \noindent
Schwinger, J. (1961). {\it Journal of Mathematical Physics,} {\bf 2}, 407. }

{\small \noindent
Tsallis, C. (1988). {\it Journal of Statistical Physics}, {\bf 52}, 479. }

{\small \noindent
Ullersma, P. (1966). {\it Physica}, {\bf 32}, 27. }

{\small \noindent
Van Kampen, N.G. (1992). {\it Stochastic Processes in Physics and Chemistry,}
North-Holland, Amsterdam. }

{\small \noindent
Weisskopf, V., and Wigner, E.P. (1930). {\it Zeitschrift f\"ur Physik}, {\bf %
63}, 54; {\bf 65}, 18. }

{\small \noindent
Wilkinson, S.R., {\it et al}. (1997). {\it Nature}, {\bf 387}, 575. }

\end{document}